\documentclass[sigconf]{acmart}

\usepackage{booktabs} 

\copyrightyear{2019}
\acmYear{2019} 
\setcopyright{iw3c2w3}
\acmConference[WWW '19]{Proceedings of the 2019 World Wide Web Conference}{May 13--17, 2019}{San Francisco, CA, USA}
\acmBooktitle{Proceedings of the 2019 World Wide Web Conference (WWW '19), May 13--17, 2019, San Francisco, CA, USA}
\acmPrice{}
\acmDOI{10.1145/3308558.3314125}
\acmISBN{978-1-4503-6674-8/19/05}
\usepackage{balance}

\makeatother \makeatletter
\newif\if@restonecol
\makeatother

\usepackage[linesnumbered,ruled,vlined]{algorithm2e}

\usepackage{graphicx}
\usepackage{color}
\usepackage{colortbl}
\usepackage{url}
\usepackage{verbatim}
\usepackage[shortlabels]{enumitem}
\usepackage{amssymb}
\usepackage{multirow}
\usepackage{balance}
\usepackage{lscape}
\usepackage{float}
\usepackage{listings}

\usepackage{url}
\usepackage{enumitem}
\usepackage{soul}

\setlist[itemize]{leftmargin=*}
\setlist[enumerate]{leftmargin=*}

\usepackage{epstopdf}

\definecolor{mygray}{gray}{.9}
\definecolor{mygreen}{rgb}{0.0, 0.5, 0.0}
\definecolor{myred}{rgb}{0.8, 0.0, 0.0}
\definecolor{mygreen1}{rgb}{0.03, 0.91, 0.87}
\definecolor{mypink}{rgb}{1.0, 0.0, 0.5}
\definecolor{mycorn}{rgb}{0.98, 0.93, 0.36}

\usepackage{dsfont}

\makeatletter
\newcommand*{\newbibstartnumber}[1]{%
  \apptocmd{\thebibliography}{%
    \global\c@NAT@ctr #1\relax
    \addtocounter{NAT@ctr}{-1}%
  }{}{}%
}
\makeatother

\begin{document}


\author{Sara Evensen, Aaron Feng, Alon Halevy, Jinfeng Li, Vivian Li, Yuliang Li, Huining Liu, \\ George Mihaila, John Morales, Natalie Nuno, Ekaterina Pavlovic, Wang-Chiew Tan,  Xiaolan Wang}
\affiliation{%
  \institution{Megagon Labs}
}
\email{{sara, aaron, alon, jinfeng, vivian, yuliang, huining, george, john, 	natalie.nuno, kate, wangchiew, xiaolan}@megagon.ai}

\renewcommand{\shortauthors}{B. Trovato et al.}

\begin{abstract}
This paper provides a sample of a \LaTeX\ document which conforms,
somewhat loosely, to the formatting guidelines for
ACM SIG Proceedings.\footnote{This is an abstract footnote}
\end{abstract}

\ccsdesc[500]{Information systems~Search interfaces}
\ccsdesc[500]{Information systems~Structured text search}
\ccsdesc[300]{Information systems~Information retrieval}
\ccsdesc[500]{Computing methodologies~Information extraction}
\ccsdesc[300]{Computing methodologies~Natural language processing}

\keywords{experiential search; subjective database; linguistic domain; information extraction;}

\title{Voyageur: An Experiential Travel Search Engine}

\newcommand{\system}{{\textsc{\small Voyageur}}}
\newcommand{\opine}{{\textsc{\small OpineDB}}}

\begin{abstract}
We describe \system, which is an application of experiential search to the domain of travel.
Unlike traditional search engines for online services, experiential search focuses on the experiential aspects of the service under consideration. In particular, \system\ needs to handle queries for subjective aspects of the service (e.g., quiet hotel, friendly staff) and combine these with objective attributes, such as price and location. \system\ also highlights interesting facts and tips about the services the user is considering to provide them with further insights into their choices.
\end{abstract}

\maketitle

\vspace{-2mm}
\section{Introduction}\label{sec:intro}

The rise of e-commerce  enables us to plan many of our future experiences with online search engines. For example, sites for searching hotels, flights, restaurants,  attractions or jobs bring a wealth of information to our fingertips, thereby giving us the ability to plan our vacations, restaurant gatherings, or  future career moves. 
Unfortunately, current search engines completely ignore any experiential
aspect of the plans they are helping you create. Instead,  they are primarily database-backed interfaces that focus on searching based on objective attributes of services, such as price range, location, or cuisine. 

\smallskip
\noindent
\textbf{The need for experiential search. }
An experiential search engine is based on the observation that at a fundamental level, users seek to satisfy an experiential need.  For example, a restaurant search is meant to fulfill a social purpose, be it romantic, work-related or a reunion with rowdy friends from college. A vacation trip plan is meant to accomplish a family or couple need such as a relaxing location with easy access to fun activities and highlights of local cuisine.
To satisfy these needs effectively, the user should be able to search on experiential attributes, such as whether a hotel is romantic or has clean rooms, or whether a restaurant has a good view of the sunset or has a quiet ambience. 

Table \ref{tab:subjective} shows a snippet from a preliminary study that investigates which attributes users care about. We asked human workers on Amazon Mechanical Turk~\cite{mturk} to provide the most 
important criteria in making their decisions in 7 common verticals. We then conservatively judge whether each criterion is an experiential one or not.
 As the table shows, the majority of attributes of interest are experiential.

\setlength{\tabcolsep}{4pt}
\begin{table}[!ht]
\vspace{-1.5mm}
\caption{Experiential attributes in different domains.}
\label{tab:subjective}
\vspace{-3mm}
\begin{tabular}{l|c|l}
\toprule
{\bf Domain}   & 
{\bf \%Exp. Attr} 
& {\bf Some examples} 
\\ \midrule
Hotel    & 69.0\% & cleanliness,  food, comfortable \\ 
Restaurant    & 64.3\% & food, ambiance, variety, service \\ 
Vacation & 82.6\% & weather, safety, culture, nightlife \\
College  & 77.4\% & 
dorm quality, faculty, diversity \\ 
Home     & 68.8\% & space, good schools, quiet, safe \\ 
Career   & 65.8\% &
work-life balance, colleagues, culture \\ 
Car      & 56.0\% & comfortable, safety, reliability  \\ \bottomrule
\end{tabular}
\vspace{-2mm}
\end{table}

To the best of our knowledge, online services for these domains today lack support for directly searching over experiential attributes.
The gap between a users' experiential needs and 
search capabilities raises an important challenge: 
{\em can we build  search systems that place the experience the user is planning at the center of the search process?}

\smallskip
\noindent 
\textbf{Challenges. }
Supporting search on experiential aspects of services is challenging for several reasons. First, the universe of experiential attributes is vast, their precise meaning is often vague, and they are expressed in text using many linguistic variations.
The experience of ``quiet hotel rooms'' can be described 
simply as ``quiet room'' or ``we enjoyed the peaceful nights'' in hotel reviews. 
Second, by definition, experiential attributes of a service are {\em subjective} and personal, and database systems do not gracefully handle such data. Third, the experiential aspect of a service may depend on how they relate to other services. For example, a significant component of a hotel experience is whether it is close to the main destinations the user plans to visit. Finally, unlike objective attributes that can be faithfully provided by the service owner, users expect that the data for experiential attributes come from other customers. Currently, such data is expressed in text in online reviews and in social media. Booking sites have made significant efforts to aggregate and surface comments from reviews. Still, while these comments are visible when a user inspects a particular hotel or restaurant, users still cannot search by these attributes.

This paper describes \system, the first search engine that explicitly models experiential attributes of services and supports searching them. We chose to build \system\  in the domain of travel because it is complex and highly experiential, but its ideas also apply to other verticals. 

The first idea underlying \system\ is that the experiential aspects of the service under consideration need to be part of the database model and visible to the user. For example, when we model a hotel, we'll also consider the time it takes to get there from the airport and nearby activities that can be done before check-in in case of an early arrival. Furthermore, \system\ will fuse information about multiple services. So the proximity of the hotel from the attractions of interest to the user is part of how the system models a hotel.  Of course, while many of the common experiential aspects can be anticipated in advance, it is impractical that we can cover them all. Hence, the second main idea in  \system\ is that its schema should be easily extensible, it should be able to handle imprecise queries, and be able to fall back on unstructured data when its database model is insufficient. 

Even with the above two principles, \system\ still faces the challenge of selecting which information to show to the user about a particular entity (e.g., hotel or attraction). Ideally, \system\ should display to the user aspects of the entity that are most relevant to the decision she is trying to make. \system\ includes algorithms for discovering items from reviews that best summarize an entity, highlight the most unique things about them, and useful actionable tips. 




\vspace{-2mm}
\section{Overview of \textsc{VOYAGEUR}} \label{sec:overview}








We illustrate the main ideas of \system. Specifically, we show how \system\ supports experiential queries and how these queries 
assist a user in selecting a hotel.

\begin{figure*}[!ht]
\vspace{-2mm}
    \centering
    \includegraphics[width=1.0\textwidth]{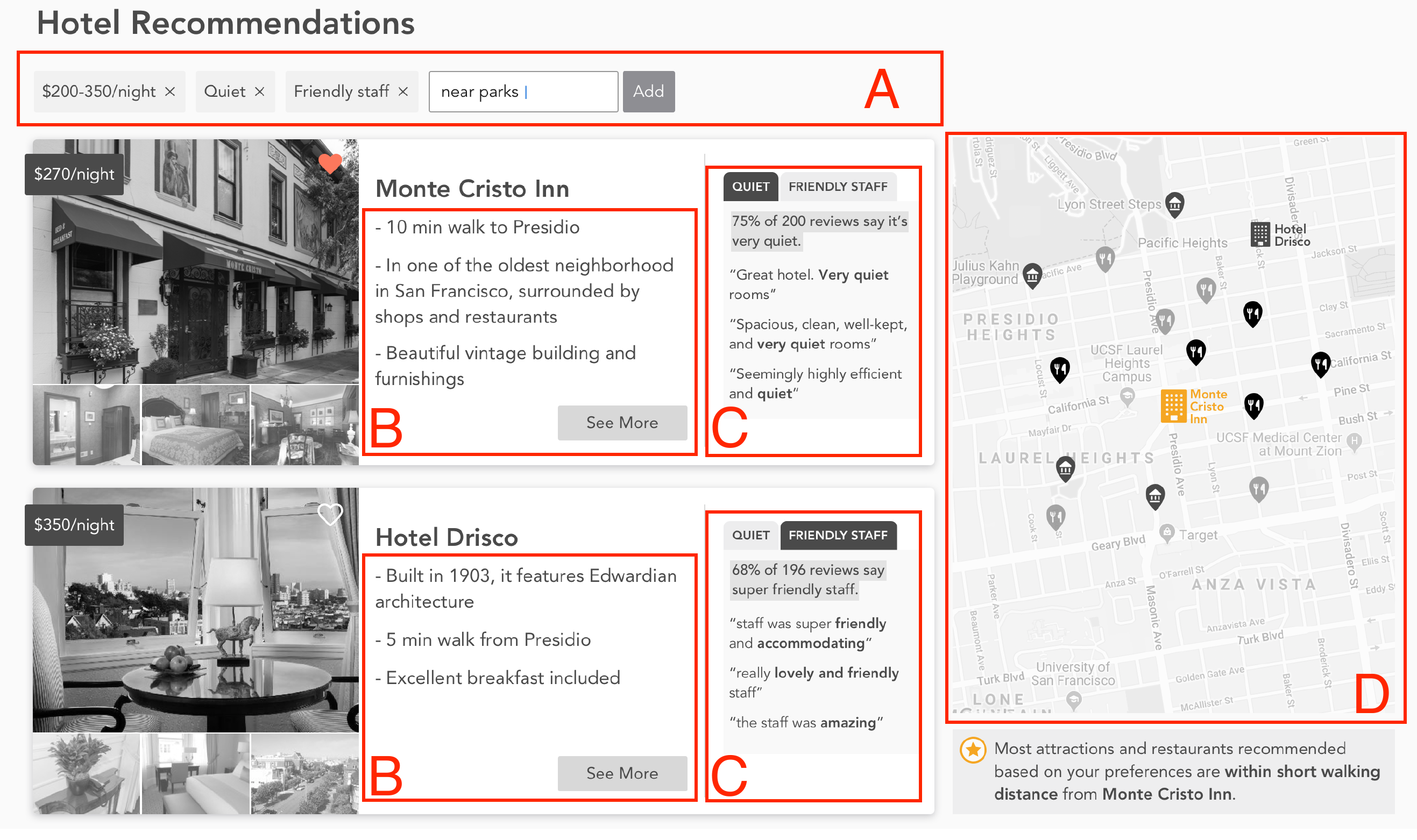}
    \vspace*{-6mm}
    \caption{A screenshot of the hotel recommendation screen of \system.
    The user can interactively customize the search (Box A), 
    to view interesting facts/tips (Box B) and a \textit{review summary} (Box C) of each recommended hotel
    and to check out a \textit{map view} (Box D) of the recommendations.}
    \label{fig:screenshot}
    \vspace{-5mm}
\end{figure*}

\smallskip
\noindent
\textbf{User scenario. }
Elle Rios is a marketing executive living in Tokyo. She is planning a vacation to \textit{San Francisco} in early October. 
Her goal is to have a relaxing experience during the vacation.
Her entire travel experience will be influenced by a variety of services, including flights,
hotels, local attractions, and restaurants. 
Elle visits the \system\ website and first 
enters the destination with the travel period. 
\system\ then displays a series of screens with recommendations for each of these services. 

In searching for hotels, Elle's experiential goal to have a relaxing stay is achieved by a balance of her {\em objective} constraints (her budget for hotels is \textit{\$250-350 per night}) and {\em subjective} criteria; she is an introvert and she knows that {\em quiet} hotels with {\em friendly staff} will help her relax\footnote{Elle can also directly search for hotels with relaxing atmosphere in \system.}.
In addition, Elle also cares about whether the hotel is conveniently located for reaching the famous and historic attractions 
she wants to visit and the high-quality vegetarian restaurants she is interested in.

Figure \ref{fig:screenshot} shows a screenshot of \system, where Elle can plan a trip that satisfies her requirements. (The screens for attraction and restaurant search are similar).
The screenshot shows how \system\ emphasizes the experiential aspects of trip planning. 
First, \system\ allows users to express their subjective criteria (in addition to their objective criteria) and generates recommendations accordingly through the \emph{experiential search} function (Box A). Second, \system\ tailors the display of {\em interesting facts and tips} and {\em summary of reviews} based on the search criteria entered (Boxes B and C).
Third, \system\ supports a series of additional features,  such as map view and travel wallet, to further improve the user's experience in the hotel search (Box D). The map view provides a holistic view of the trip by putting together all recommended entities of different types. The travel wallet, as we explain later, takes into consideration the user's travel history and preferences if the user chooses to share them with \system.

\smallskip
\noindent
\textbf{Experiential search. }
Elle expresses her objective and experiential/subjective 
criteria as 
query predicates  to \system's interface.
While the objective requirements like ``\$250 to 300 per night'' can be directly modeled and queried in 
a typical hotel database,
answering predicates like \emph{``quiet''} and \emph{``friendly staff''} is challenging as these are subjective terms and cannot be immediately modeled
in a traditional database system.
\system\ addresses this challenge with a subjective database engine
that explicitly models the subjective attributes and answers subjective query predicates.
\system\ extracts subjective attributes
such as {\it room quietness} and {\it staff quality} from hotel customer reviews, builds a summary of the variations of these terms, and
then matches those attributes with the input query predicates.


In Figure \ref{fig:screenshot}, \system\ generates a ranked list of hotels by matching the query predicates specified by Elle
with the subjective attributes extracted from the underlying hotel reviews.
The review summaries (Box C) show that the selected hotels are clearly good matches. 
Specifically, \system\ recommends Monte Cristo since 75\% of 200 reviewers agree that it is very quiet and it has friendly staff (not shown).
Hotel Drisco, next on the list, is recommended because 68\% of 196 reviewers agree that it has friendly staff and the it is also very quiet (not shown).

\smallskip
\noindent
\textbf{Interesting facts and tips. }
Along with the search results, \system\ shows snippets of travel tips and/or interesting facts of each result (Box B) it thinks is relevant for Elle. 
An interesting fact typically highlights an unusual or unique experience 
about the service.
For example, being very close to  Presidio Park (one of the largest parks in San Francisco) is unique to Monte Cristo Inn
and Hotel Drisco and is thus an interesting fact to show for each hotel. The fact that
Monte Cristo Inn is a ``beautiful vintage building and furnishings" is unique only to 
Monte Cristo Inn.
Such interesting facts can be important for decision making.
It also enables Elle to better anticipate the type of experiences
that will be encountered at the hotel~\cite{anticipation}.
On the other hand, tips are snippets of information that propose
a potential action  the user  may take to either avoid a negative experience
or create a positive one~\cite{tipmining}. For example, a useful tip for a hotel may be that there is free parking two blocks away.


While the interesting facts and tips are useful, they are not always available for every service and 
can be incomplete. Existing work \cite{suggestion,tipmining} proposed mining useful travel tips from customer reviews with promising results.
In \system, we formulate the problem of finding tips and interesting facts as a query-sentence matching problem to find tips and interesting facts relevant to the users' query.
Our algorithms prefer to select sentence snippets from reviews
to match the user's query.
The challenge we face is that 
users' query predicates and 
the tips/facts in reviews are described in different vocabularies and linguistic forms.
Moreover, labeled data for the matching task is generally not available. Thus,
novel techniques are needed to construct good matching functions between queries and
sentence snippets.

\smallskip
\noindent
\textbf{Review summarization. } 
The summary of reviews (Box C) provides Elle with an explanation of why the specific hotel is recommended and 
the summary saves her from reading the repetitive and lengthy reviews.
\system\ summarizes the reviews of each recommended hotel in two different formats:
(1) statistical statements and (2) sample review snippets.
For example, in Figure \ref{fig:screenshot}, \system\ summarizes the room quietness attribute of Monte Cristo Inn
with the statistical statement ``75\% of 200 reviews say it is very quiet'' and 
3 randomly selected sample review snippets that match the \emph{quiet} requirement.



\smallskip
\noindent
\textbf{Additional features. } The following features further improve Elle's ability to create a positive travel experience: 

\begin{itemize}\parskip=0pt
\item \textbf{Map view. } In each recommendation screen, a map view (Box D of Figure \ref{fig:screenshot})
marks the locations of the recommended hotels.
Whenever a hotel is selected, the map view is centered at a chosen hotel and 
shows the recommended local attractions and restaurants so
the user can better plan how to travel between these places.

\item \textbf{Travel wallet. } Users have the option of creating a 
\emph{travel wallet}, which is similar to the Wallet feature on many smartphones. It contains information about the user that she shares only when she chooses to. In the case of a travel wallet, this information records her travel preferences. The travel wallet can be created explicitly by answering questions or can be collected automatically from previous travels.
The travel wallet is used by \system\ to further personalize the search results.

\item \textbf{Trip summary. }
Finally, after making several choices (flight, hotel, attractions, etc.)
through all the recommendation screens, the user can view
a summary of the trip underlying the key experiential components.
In Elle's case, the summary includes a timeline with important dates,
transportation methods to/from the airport, and
tips/facts about the chosen hotel and each planned tourist attractions.

\end{itemize}

\vspace{-2.5mm}
\section{Implementation of \system} \label{sec:implementation}

We briefly touch upon the technology underlying
 \system.

\vspace{-2mm}
\subsection{A subjective database engine}

As mentioned, the main challenge of building a successful experiential search engine
is the modeling and querying of the \emph{subjective attributes}, where there is typically no ground truth to the values of such attributes. Examples of such attributes include the cleanliness of hotel rooms, quality of the food served, 
and cultural value of tourist attractions.
They are not explicitly modeled in today's search engines and therefore not directly queryable. 

\system\ is developed on top of \opine~\cite{opinedb19corr}, a \emph{subjective database engine}.
\opine\ goes beyond traditional database engines by supporting
the modeling, extraction, aggregating, and effective query processing of 
the subjective data. 
Next, we illustrate the key design elements of \opine\ by showcasing 
its application to hotel search in \system.

\smallskip
\noindent
\textbf{Data model, extraction, and aggregation. }
The main challenge in modeling subjective attributes is the wide range of linguistic variations
with which the attributes are described in text. Consider an attribute {\tt room\_quietness} of hotels. 
The review text can be of various forms such as 
(1) ``{\em the neighborhood seems very quiet at night}'',
(2) ``{\em on busy street with traffic noise}'', or
(3) ``{\em quiet and peaceful location'}'.
In addition, \opine\ needs to {\em aggregate} these phrases into a meaningful signal for answering queries,
which may themselves include new linguistic variations.

\opine\ models subjective attributes with a new data type called the {\em linguistic domain} and provides an aggregated view of the linguistic domain through a {\em marker summary}.
The linguistic domain of an attribute contains
all phrases for describing the attribute from the reviews. E.g., ``quiet at night'', ``traffic noise'', and ``peaceful location''.
A subset of phrases is then chosen as
the {\em domain markers} (or {\em markers} for short) for each linguistic domain.
The phrases are aggregated based on the markers to constitute the marker summary.

For quietness, the markers might be $[$very\_quiet, average, noisy, very\_noisy$]$.
To construct the marker summary of a hotel's quietness, \opine\ 
 needs to assign each quietness phrase to its closest marker and
compute the frequencies of the markers. 
For example, the summary $[$very\_quiet:20, average: 70, noisy:30, very\_noisy:10$]$ 
for a hotel would represent that the hotel is closer to being {\tt average} in quietness than to the other markers. 

The linguistic domain is obtained by extracting phrases from reviews.  
Various techniques are available for this task in opinion mining and sentiment analysis \cite{liu2012sentiment, absa}.
The marker summaries are currently histograms computed from the extraction relations. However,
we can also leverage more complex aggregate functions.




\smallskip
\noindent
\textbf{Query processing. } The query predicates
from Box A is formulated as an SQL-like query
for \opine\ to process.
\vspace*{1mm}

\begin{tabular}{p{1cm}p{3in}}
{\bf select} &  * \hspace{0.3cm}{\bf from} {\tt Hotels} $h$\\
{\bf where}  & $\mathtt{price\_pn} \leq 350$ {\bf and} $\mathtt{price\_pn} \geq 200$ {\bf and} \\
             & ``{\em quiet}'' {\bf and} ``{\em friendly staff}''
\vspace*{1mm}
\end{tabular}
Here, price\_pn is an objective attribute of the Hotels relation while ``{\em quiet}'' and ``{\em friendly staff}'' are subjective predicates.
\opine\ needs to interpret these predicates using the linguistic domains in order to find the best subjective attributes of the Hotels relation that can be used to answer them.
In general, this is not a trivial matching problem since the query terms may not be directly modeled in the schema.  
For example, the user may ask for ``romantic hotels'', but the attribute for romance might not be in the schema.
For such cases, \opine\ leverages a combination of NLP and IR techniques to find a best-effort reformulation of the query term into a combination of schema attributes. For example, for  ``romantic hotels'', \opine\ will match it to a combination of ``exceptional service'' and ``luxurious bathrooms''
which are modeled by the schema.

After computing the interpretation, \opine\ uses the marker summaries to compute
a {\em membership score} for each pair of hotel and query predicate. Finally, \opine\ combines  multiple predicates using a variant of fuzzy logic.


\vspace{-2mm}
\subsection{Mining interesting facts and tips }

We formulate the problem of finding useful travel tips and interesting facts as
a query-sentence matching problem. We adopt an approach similar to an existing
work for mining travel tips from reviews \cite{tipmining}.
The approach consists of a \emph{filtering phase} and
a \emph{ranking phase}. 

\smallskip
\noindent
\textbf{Filtering phase. } This phase constructs a set of candidate
tip/fact sentences by applying filters and classifiers to all the review sentences.
According to \cite{tipmining}, effective filters for tips include phrase patterns
(e.g, sentences containing ``make sure to'') and part-of-speech tags (e.g, sentences starting with verbs).
For constructing the candidate set of interesting facts, we select sentences that contain a least one 
\emph{informative token}, which are words or short phrases frequently mentioned in reviews of the target entity
but not frequently mentioned in reviews of similar entities. 
We also found that an interesting fact is more likely to appear in sentences with an extreme sentiment (very positive or negative).
So we also apply sentiment analysis to select such sentences.
For both tips and interesting facts, we further refine the candidate sets by removing duplicates, i.e.,
sentences of similar meaning or the unimportant ones. We do so by applying TextRank \cite{textrank}, a classic algorithm
for sentence summarization.

\smallskip
\noindent
\textbf{Ranking phase. } Instead of simply selecting candidates for interesting facts/tips,
sets, we implemented a novel ranking function for finding
candidates that best match the user's query predicates.
The ranking function considers not only the \emph{significance} of a candidate
as computed in the filtering phase but also the \emph{relevance} of the candidate
with the query. Measuring the relevance is not trivial since the tips/interesting facts
can use vocabularies different from the ones used in the query.
In the previous example, a fact that matches the query ``near park'' is ``10 min walk to Presidio''
which has no exact-matched word. The similarity function leverages a combination of NLP and IR techniques, 
analogous to query interpretation in \opine.



\vspace{-2mm}
\subsection{Datasets and tools}
The \system\ demo will serve hotels, attractions, and restaurants search in the San Francisco area in a browser-based web app.
We collected the underlying data from two datasets: the Google Local Reviews\footnote{\url{http://cseweb.ucsd.edu/~jmcauley/datasets.html\#google\_local}}
and the TripAdvisor datasets\footnote{\url{http://nemis.isti.cnr.it/~marcheggiani/datasets/}}.   
Our dataset consists of 18,500 reviews of 227 hotels, 6,256 reviews of 545 attractions and 67,382 reviews of 4,007 restaurants.
We implemented the front end of \system\ using the JavaScript library React.
\vspace{-2mm}
\section{Conclusion} \label{sec:conclude}
The motivation for \system\  is based on  the discrepancy between the needs of users searching for services and the current state of search engines. The ideas of \system\ are applicable to many other verticals beyond travel.  At the core of the technical challenges that \system\ and systems like it need to address is the ability to discover and aggregate evidence from textual reviews in response to user queries.
This is a technical challenge that draws upon techniques from NLP, IR and Database technologies.

\vspace{-1mm}

\bibliographystyle{ACM-Reference-Format}
\bibliography{references}

\balance 

\end{document}